\documentclass[letterpaper]{article} 
\usepackage[preprint]{aaai2027}  
\usepackage[hyphens]{url}  
\usepackage{graphicx} 
\usepackage{natbib}  
\usepackage{caption} 
\usepackage{algorithm}
\usepackage{algorithmic}
\usepackage{amsmath}
\usepackage{multirow}
\usepackage{amssymb}
\usepackage{booktabs}
\usepackage[table]{xcolor}
\definecolor{oursgray}{gray}{0.88}
\usepackage{newfloat}
\usepackage{listings}
\DeclareCaptionStyle{ruled}{labelfont=normalfont,labelsep=colon,strut=off} 
\floatstyle{ruled}
\newfloat{listing}{tb}{lst}{}
\floatname{listing}{Listing}

\usepackage{booktabs}

\title{DREvo: Distilling Recalibrated Historical Experience for Harness Self-Evolution}

\author{
    Hanghui Guo\textsuperscript{\rm 1}, Weijie Shi\textsuperscript{\rm 2}, Zhangze Chen\textsuperscript{\rm 3}, Shengxiang Xu\textsuperscript{\rm 1}, Yishu Wang\textsuperscript{\rm 1}, Yimei Zhang\textsuperscript{\rm 4}, Wangze Ni\textsuperscript{\rm 5}, Jia Zhu\textsuperscript{\rm 3}, Shimin Di\textsuperscript{\rm 1}
}
\affiliations{
    \textsuperscript{\rm 1}School of Computer Science and Engineering, Southeast University, China\\
    \textsuperscript{\rm 2}Department of Computer Science and Engineering, Hong Kong University of Science and Technology, China\\
    \textsuperscript{\rm 3}Zhejiang Key Laboratory of Intelligent Education Technology and Application, Zhejiang Normal University, China\\
    \textsuperscript{\rm 4}College of Computer Science and Technology, Zhejiang University of Technology, China\\
    \textsuperscript{\rm 5}College of Computer Science and Technology, Zhejiang University, China\\

}

\begin{document}

\maketitle

\begin{abstract}
Harness plays a critical role in large language model agent performance, and building a high-performing harness requires substantial expert effort. Therefore, recent research has increasingly explored harness self-evolution, which iteratively proposes, evaluates, and improves harnesses using historical trial experience. However, accumulated historical experience does not always translate into stable search guidance, and performance often fluctuates substantially across evolution iterations, making it difficult to reliably discover high-performing harnesses under a limited evolution budget. We identify two limitations in how existing harness self-evolution methods leverage historical experience: \textit{(1) Lack of dynamic reassessment of whether historical experience remains valid for the current harness}, and \textit{(2) Lack of explicit mechanisms for translating valid historical experience into actionable search directions}. To address these limitations, we propose a new harness self-evolution method, named
\textbf{DREvo}, which integrates function-level evidence anchoring, state-dependent evidence recalibration, and role-conditioned search intent distillation to determine \textit{which historical evidence remains valid} and \textit{where the harness should evolve next}.
Under limited evolution budgets, DREvo exhibits smoother evolution trajectories, achieves the highest accuracy on all five benchmarks, and delivers average gains of 16.2\% and 13.8\% over the evaluated baselines on domain reasoning and agentic tasks, respectively.


\end{abstract}


\section{Introduction}

The capabilities of a large language model (LLM) agent are shaped not only by its underlying foundation model, but also by its external execution framework, commonly referred to as a \textit{harness}~\cite{rajasekaran2026harness, lopopolo2026harness, chen2026unlocking}.

The harness specifies how an agent operates during execution, including how it constructs context, manages memory, invokes tools, and parses output, and can therefore substantially affect its behavior~\cite{yang2024swe, li2024embodied, liu2026toolscope}. However, effective harness design still relies heavily on human expertise. Developers must inspect evaluation failures and execution traces, diagnose interactions among components, and repeatedly coordinate revisions across the entire harness~\cite{guo2025dior, shen2026acr, hu2025automated}.
As tasks grow more complex and harnesses involve more components, manual refinement becomes increasingly labor-intensive and difficult to scale~\cite{zhang2025aflow, qiao2025benchmarking, yang2024large}, motivating recent research into \textit{harness self-evolution}.

Existing harness self-evolution methods, such as Meta-harness~\cite{lee2026meta}, AHE~\cite{lin2026agentic}, and ACE~\cite{zhang2025agentic}, typically follow a \textit{history-driven paradigm}. Specifically, at each iteration, the candidate harness is executed and evaluated, and the resulting records or their processed summaries are accumulated as historical experience, including candidate code, execution trajectories such as model responses and tool interactions, and derived feedback such as failure summaries. This historical experience is then provided to the proposer in subsequent iterations to guide the generation of new candidate harnesses. Overall, this pipeline aims to optimize harness designs through iterative \textit{propose-evaluate-feedback} cycles~\cite{agrawal2025gepa}.

\begin{figure}
    \centering
    \includegraphics[width=\linewidth]{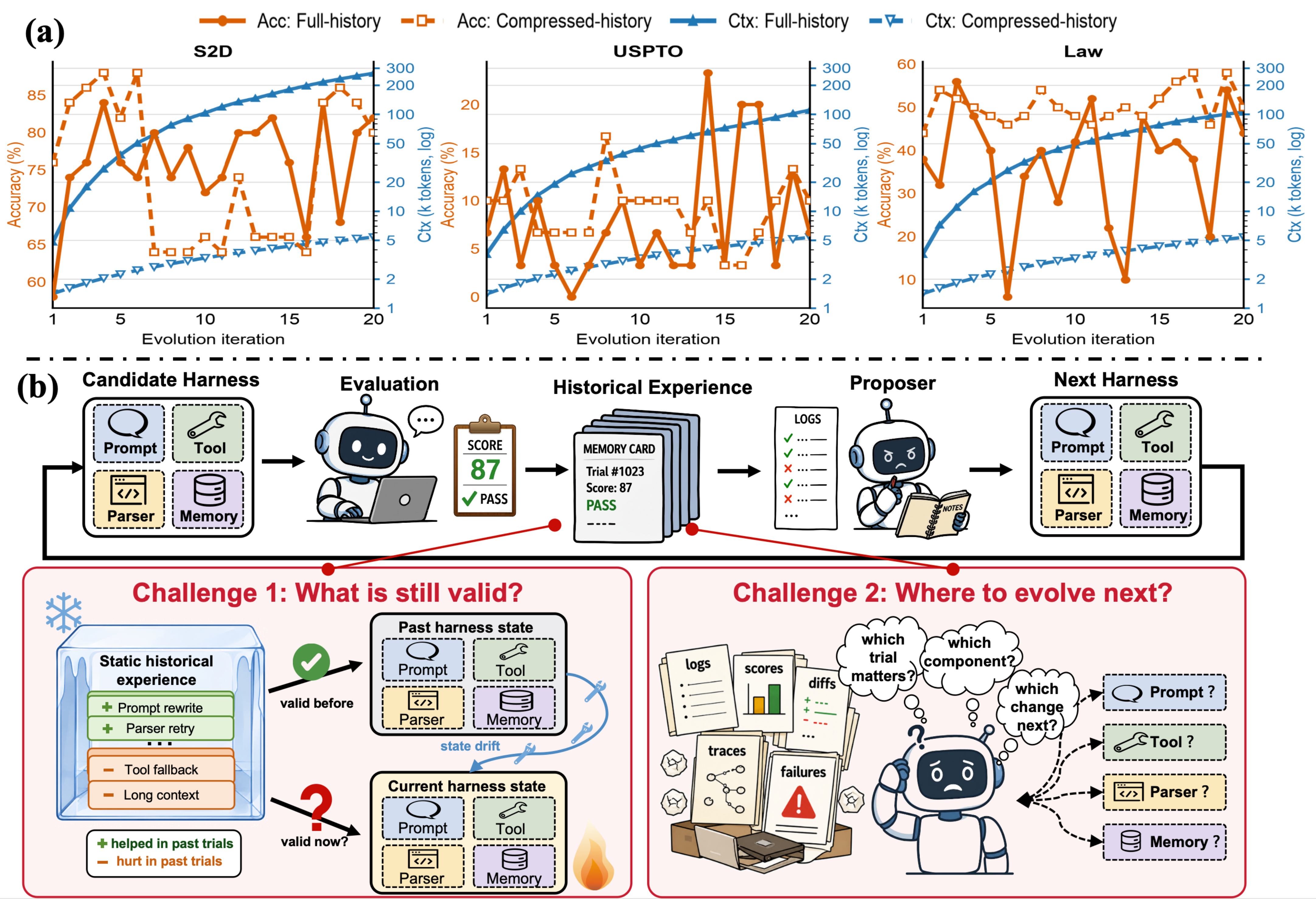}
   \caption{ 
(a) Full-history and compressed-history reuse \textit{(Meta-harness, Claude Opus 4.6, 1M-token context window)} both exhibit non-stationary evolution across the S2D, USPTO, and Law datasets, with accuracy repeatedly regressing and recovering across iterations despite accumulating historical experience. (b) Two limitations in harness self-evolution: \textbf{What is still valid?} and \textbf{Where to evolve next?}
    }
    \label{fig:motivation}
\end{figure}


In practice, this process requires generating and evaluating a candidate harness at every iteration. Each iteration involves multiple LLM calls and task-environment interactions, often resulting in long rollout times, substantial API and computational costs. However, as shown in Figure~\ref{fig:motivation}(a), 
under both full-history and compressed-history reuse, accuracy does not exhibit a stable upward trend as historical experience accumulates, but instead fluctuates markedly across iterations, revealing pronounced non-stationarity in the evolution trajectory. This suggests that the continued accumulation of historical experience does not necessarily provide effective search guidance for subsequent evolution. Consequently, many iterations that fail to produce a better harness continue to consume evaluation budget. Under a limited evolution budget, history-driven harness self-evolution may therefore fail to reliably discover a high-performing harness.

To further investigate this issue, we identify two key limitations in how existing harness self-evolution methods leverage historical experience, as summarized in Figure~\ref{fig:motivation}(b):

\textbf{(1) What is still valid?}
Historical experience in harness self-evolution is highly state-dependent and cannot be reused unconditionally. Each trial record reflects the effect of a component modification under a specific harness implementation, execution outcome, and task context. As the evolution process proceeds, subsequent modifications may alter the component state on which earlier experience depends. As a result, a previously effective modification may become invalid under the current state, while a previously harmful or ineffective direction may become viable again due to state changes. Existing methods typically preserve such experience as statically valid evidence, but lack explicit calibration of its applicability to the current harness state. Thus, the proposer may base its updates on mismatched evidence, causing stale experience to distort the subsequent search trajectory.


\textbf{(2) Where to evolve next?}
Historical experience in harness self-evolution is not merely a record of past trials, but evidence for guiding subsequent harness modifications. However, turning such evidence into effective updates requires deciding which harness component to modify and how to modify it.
A harness typically contains multiple interrelated components and substantial code, while historical experience contains heterogeneous evidence. Simply exposing the proposer to full historical experience or summaries leaves these decisions implicit. This may produce ineffective evolution directions and make it more difficult to discover a high-performing harness under a limited budget.
Thus, harness self-evolution requires an explicit update policy that balances exploration and exploitation across iterations, enabling past experience to stably guide subsequent search.

To address these limitations, we introduce a novel harness self-evolution method, named \textbf{DREvo}. Specifically, DREvo consists of three main components:
\textbf{(1) Function-Level Evidence Anchoring.}
It transforms coarse iteration-level trial logs into function-anchored evidence units, making heterogeneous feedback localizable and traceable to specific harness components while providing the basis for subsequent evidence recalibration and search-intent distillation.
\textbf{(2) State-Dependent Evidence Recalibration.}
It recalibrates each evidence unit according to its reliability across prior uses and its structural compatibility with the current harness, thereby reassessing the validity of historical experience under the evolving harness state.
\textbf{(3) Role-Conditioned Search Intent Distillation.}
It retrieves historical evidence relevant to failure patterns observed in the current iteration and distills it into an explicit, role-conditioned search intent that specifies where and how the harness should evolve next.

The main contributions of our work are as follows:
\begin{itemize}
    \item We identify \textit{substantial performance fluctuations} as a key reliability challenge in harness self-evolution and trace them to two limitations concerning \textit{what is still valid} and \textit{where to evolve next}, both of which reduce search effectiveness and hinder the reliable discovery of high-performing harnesses under limited evolution budgets.

    \item We propose \textbf{DREvo}, a novel harness self-evolution method that transforms raw trial logs into fine-grained, function-anchored evidence units, dynamically recalibrates their validity under the current harness state, and distills the calibrated evidence into an explicit search intent through evidence-conditioned role selection.

    \item We validate the effectiveness of DREvo through extensive experiments, showing smoother evolution trajectories with fewer performance oscillations and average gains of 16.2\% and 14.2\% over the evaluated baselines on domain reasoning and agentic tasks, respectively.
\end{itemize}

\section{Related Work}

\noindent
\textbf{Harness Engineering.}
Harness design plays a crucial role in shaping the capabilities of large language model (LLM) agents~\cite{ning2026code,meng2026agent,wang2026openclaw}. For example, SWE-agent provides interfaces for code navigation, editing, and testing~\cite{yang2024swe}. OpenHands integrates code execution, terminal interaction, and sandbox environments~\cite{wang2025openhands}. Mini-SWE-Agent adopts a simplified Bash interface with a linear message history~\cite{yang2024swe}. Terminus-2 allows the model to directly operate a persistent terminal session~\cite{merrill2026terminal}. Terminus-KIRA employs native tool calling for terminal tasks~\cite{krafton2026terminus}. Claude Code integrates codebase understanding, file editing, and command execution into a project development workflow. Collectively, these systems demonstrate how harness design shapes agent--environment interaction and task execution, but their architectures and interfaces remain primarily hand-engineered.

\begin{figure*}
    \centering
    \includegraphics[width=\linewidth]{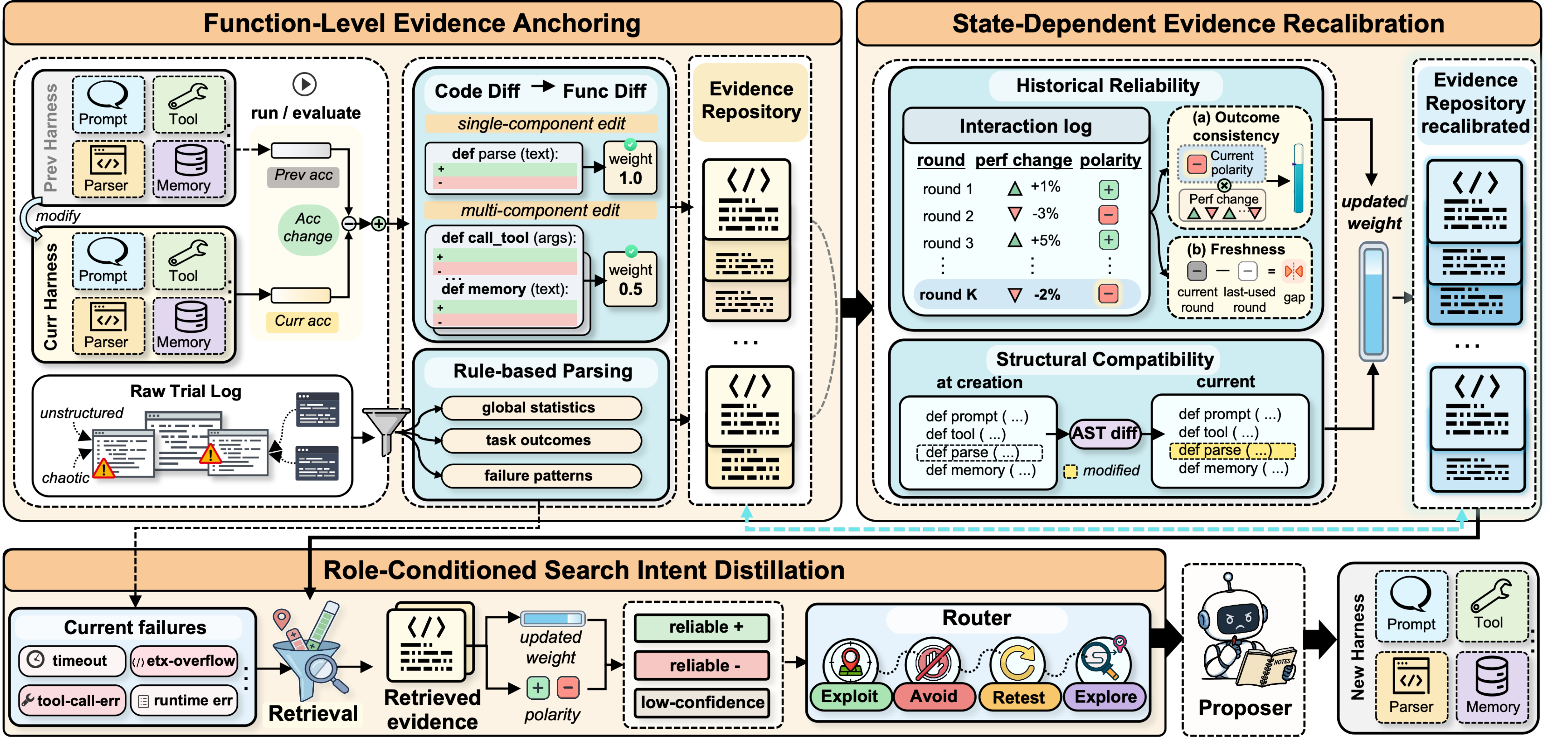}
    \caption{Detailed overview of the DREvo framework, which anchors historical evidence at the function level, recalibrates its validity under the current harness state, and distills it into explicit guidance for harness self-evolution under limited budgets.}
    \label{fig:method}
\end{figure*}

\noindent
\textbf{Harness Self-Evolution.}
Recent advances in LLMs have improved their code generation abilities~\cite{liang2026rsda}. This progress enables LLMs to drive harness self-evolution, reducing reliance on manual engineering. Some work focuses primarily on context management within the harness. ACE uses LLMs to build an evolving playbook that accumulates and revises task strategies~\cite{zhang2025agentic}. MCE uses LLM agents at two optimization levels to jointly evolve context engineering skills and context artifacts~\cite{ye2026meta}. However, these methods optimize only the context component of the harness and may therefore be less effective in complex tasks involving multiple interacting harness components. Later studies extend optimization to a wider range of harness components. AutoHarness synthesizes task-specific constraint code from environmental feedback~\cite{lou2026autoharness}. Self-Harness examines how harness modifications should be adapted to differing requirements of foundation models~\cite{zhang2026self}. Meta-harness guides harness generation by reusing compressed summaries of candidate code, evaluation scores, and execution traces~\cite{lee2026meta}. AHE generates improved harnesses through trajectory distillation and verifiable feedback from execution outcomes~\cite{lin2026agentic}. These studies broaden the scope of harness optimization and advance harness self-evolution through execution feedback and modification validation.



However, prior studies primarily focus on constructing end-to-end pipelines for harness self-evolution, while the reliability of history-driven search within this process remains underexplored. Specifically, these methods typically reuse records or summaries as fixed context. As the harness state changes, historical experience may become stale. Moreover, search decisions often rely on the proposer’s implicit reasoning, leading to unstable search directions during evolution. These issues motivate DREvo, which anchors, recalibrates, and distills historical evidence into explicit search intent.



\section{Proposed Approach} 

\subsection{Problem Setup}

\noindent
\textbf{Harness Self-Evolution.}
We first formalize the conventional history-driven harness self-evolution paradigm. Given an initial harness $h_0$, an evaluation environment $\mathcal{E}$, and $T$ evolution iterations, harness self-evolution performs history-conditioned sequential optimization. Before evolution, $h_0$ is evaluated under $\mathcal{E}$ to obtain the initial trial log $\tau_0$ and initialize the experience repository $\mathcal{R}_0$. At iteration $t$ ($1\leq t\leq T$), the proposer $\mathcal{P}$ first generates a candidate harness $h_t$ conditioned on the preceding harness $h_{t-1}$ and repository $\mathcal{R}_{t-1}$. Evaluating $h_t$ under $\mathcal{E}$ then produces a trial log $\tau_t$, which is appended to $\mathcal{R}_t$. The objective is to select the best-performing harness:
\begin{equation}
h_t\sim\mathcal{P}(\cdot\mid h_{t-1},\mathcal{R}_{t-1}),
\qquad
h^\star=\arg\max_{h\in\mathcal{H}}J_{\mathcal{E}}(h),
\end{equation}
where $\mathcal{H}=\{h_0,h_1,\ldots,h_T\}$ contains the initial harness and all candidates generated during evolution, and $J_{\mathcal{E}}$ denotes their evaluation performance.

\noindent
\textbf{Overview of DREvo.}
In the conventional process above, the proposer directly conditions each new harness on accumulated experience. However, as the harness evolves, historical evidence may become invalid and does not explicitly indicate which component to modify or how to modify it. These limitations can destabilize the search and hinder the reliable discovery of a high-performing harness within \(T\) iterations. To address them, we propose \textbf{DREvo}, as illustrated in Figure~\ref{fig:method}. DREvo comprises three components: \textit{Function-Level Evidence Anchoring}, \textit{State-Dependent Evidence Recalibration}, and \textit{Role-Conditioned Search Intent Distillation}.

\subsection{Function-Level Evidence Anchoring}

Raw trial logs are organized at the iteration level, mixing harness modifications, execution trajectories, and evaluation feedback. When directly using these logs, the proposer $\mathcal{P}$ must infer the relationships between component modifications and execution outcomes from a growing volume of heterogeneous information, making reusable experience difficult to extract. To address this issue, we introduce \textit{Function-Level Evidence Anchoring}, which uses functions as basic anchoring units to associate each modification with its observed feedback, thereby localizing the resulting evidence to specific harness components and storing it in $\mathcal{R}_t$.

Specifically, we represent a harness as a set of functional components \(\mathcal{C}\), such as tool use and memory management. Each component \(c\in\mathcal{C}\) is represented by its corresponding function implementation \(f_c\), thereby establishing an explicit anchor between the component and its evolving code state.

At each iteration, DREvo first parses the structured trial log \(\tau_t\) to extract feedback \(\mathbf{o}_t=(\mathbf{g}_t,\mathcal{Q}_t,\mathcal{F}_t)\). Here, \(\mathbf{g}_t\) records global statistics, including the pass rate, average score, and resource consumption; \(\mathcal{Q}_t\) contains task-level outcomes, including the pass status, score change, and failure summary of each task; and \(\mathcal{F}_t\) summarizes the frequencies of different failure patterns, such as timeouts, runtime exceptions, tool-call errors, and context overflow.

DREvo then computes a line-level code diff between $h_{t-1}$ and $h_t$. Each diff hunk is mapped to its enclosing functional component according to the source-span boundaries of $\{f_c\}_{c\in\mathcal{C}}$, allowing DREvo to identify the modified components and collect their corresponding code changes:
\begin{equation}
\operatorname{FuncDiff}(h_{t-1},h_t)
=
\left\{
(c,\delta_{t,c})
\mid c\in\mathcal{C}_t^{\Delta}
\right\},
\end{equation}
where $\mathcal{C}_t^{\Delta}$ is the set of modified components, $c$ identifies a specific component, and $\delta_{t,c}$ records the code changes made to the function corresponding to $c$. In practice, DREvo encourages the proposer to modify a single target component whenever possible, so that the resulting feedback can be more directly attributed to a component-scoped code change. 

For each modified component \(c\in\mathcal{C}_t^{\Delta}\), DREvo further determines two attributes of the corresponding evidence unit: an initial binary polarity \(\rho_e\) and weight \(w_e\). To determine the polarity, DREvo computes the trial-level performance change \(\Delta_t=J_{\mathcal{E}}(h_t)-J_{\mathcal{E}}(h_{t-1})\), setting \(\rho_e=-1\) if \(\Delta_t<0\) and \(\rho_e=+1\) otherwise, including ties. By default, \(w_e=1.0\) for a single-component edit and \(w_e=0.5\) for a multi-component edit, reflecting lower attribution confidence because component-specific effects cannot be isolated in the latter case. DREvo then constructs the function-anchored evidence unit \(e_{t,c}=(c,f_c,\delta_{t,c},\mathbf{o}_t,\Delta_t,\rho_e,w_e,p_t)\), where \(p_t\) points to the provenance traces in \(\tau_t\).

All constructed evidence units $e_{t,c}$ are appended to $\mathcal{R}_t$. When an evidence unit is subsequently reused to guide evolution, its polarity is updated using the same rule based on the resulting performance change, while its weight is recalibrated through \textit{State-Dependent Evidence Recalibration}.

\subsection{State-Dependent Evidence Recalibration}

Function-Level Evidence Anchoring transforms trial logs into function-level evidence, but each piece of evidence is derived under a specific harness state. As the harness evolves, previously effective experience may become obsolete, whereas failed directions may become viable again. Thus, historical evidence must be re-evaluated before guiding subsequent evolution. To address this issue, we introduce \textit{State-Dependent Evidence Recalibration} to dynamically calibrate evidence validity under the current harness state.
Specifically, the validity of historical evidence is assessed from two perspectives: (1) whether the evidence remains reliable based on prior uses (historical reliability $r_t(e)$), (2) whether its associated component state remains compatible with the current harness (structural compatibility $q_t(e)$).

To estimate historical reliability, DREvo maintains an interaction log for each evidence unit. The log is initialized with the performance change and outcome polarity observed when the evidence is created, and subsequently records the corresponding outcome of each invocation. 
Based on this accumulated interaction history, DREvo characterizes historical reliability from two complementary signals: outcome consistency and freshness, capturing whether previous uses remain supportive and temporally relevant during evolution. 

Outcome consistency measures how strongly the recorded outcomes support the current evidence polarity, weighted by their performance changes. Let $\mathcal{I}_t(e)$ denote the set containing the creation record and all subsequent invocation records of evidence $e$ up to iteration $t$, where $\rho_j(e)$ and $\Delta_j(e)$ denote the outcome polarity and performance change recorded at iteration $j$, respectively. We define it as $\operatorname{cons}_t(e)$:
\begin{equation}
\operatorname{cons}_t(e)=
\frac{
\sum_{j\in\mathcal{I}_t(e)}
\mathbb{I}[\rho_j(e)=\rho_e]\,
|\Delta_j(e)|
}{
\sum_{j\in\mathcal{I}_t(e)}
|\Delta_j(e)| + \epsilon
},
\end{equation}
where $\rho_e$ denotes the current evidence polarity, $\mathbb{I}[\cdot]$ is the indicator function, and $\epsilon$ is a small constant that prevents a zero denominator.
Thus, $\operatorname{cons}_t(e)$ represents the fraction of the total observed performance variation that supports the current evidence polarity. It becomes high when more prior invocations, particularly those with large performance changes, agree with the current polarity; otherwise, the value tends to be low. Therefore, $\operatorname{cons}_t(e)$ reflects the reliability of the current polarity under historical usage.

Freshness measures temporal relevance with a fixed decay:
\begin{equation}
\operatorname{fresh}_t(e)=
0.9^{\,t-\ell_t(e)},
\end{equation}
where $\ell_t(e)$ denotes the iteration of the most recent record in $\mathcal{I}_t(e)$, corresponding to either its creation or latest invocation. Freshness therefore decays as the evidence remains unused. 

Since historically reliable evidence should be both outcome-consistent and recent, DREvo defines its historical reliability as $r_t(e)=\operatorname{cons}_t(e)\operatorname{fresh}_t(e)$.


Beyond historical reliability, DREvo estimates structural compatibility. Each evidence unit is tied to the component implementation under which it was created; as the harness evolves, that implementation may change and reduce the applicability of the evidence. DREvo therefore checks whether the evidence-anchored component remains structurally compatible with its current state before reuse.

Specifically, DREvo retrieves the component code associated with the evidence and locates the corresponding component in the current harness through function-level anchoring. It then parses both component implementations into abstract syntax trees (ASTs), which serve as a \textit{lightweight structural proxy for compatibility rather than a proof of semantic equivalence}, and normalizes superficial lexical details, such as identifiers, while preserving behavior-sensitive literal values, prompt strings, control-flow structures, call patterns, and state-update operations.
Let $T_e$ and $T_t$ denote the AST of the component at evidence creation and in the current harness, respectively. The structural compatibility of the component state is defined as $q_t(e)$:
\begin{equation}
q_t(e)=
1-
\frac{
\operatorname{TED}(T_e,T_t)
}{
|T_e|+|T_t|
},
\end{equation}
where $\operatorname{TED}(T_e,T_t)$ denotes the minimum number of node insertions, deletions, and substitutions required to transform $T_e$ into $T_t$, and $|T_e|$ and $|T_t|$ denote their respective numbers of AST nodes.
A smaller tree edit distance indicates that the control structures, function calls, and state-update logic underlying the evidence are better preserved in the current component, resulting in higher state compatibility. Conversely, substantial structural changes reduce $q_t(e)$. Because comparison is restricted to the anchored component, unrelated changes do not affect compatibility.
Finally, DREvo computes the calibrated evidence weight as:
\begin{equation}
w_e=\frac{2r_t(e)q_t(e)}{r_t(e)+q_t(e)+\epsilon},
\end{equation}
where $w_e\in[0,1]$. This aggregation treats reliability and compatibility as jointly necessary: a high value in one factor cannot fully compensate for a low value in the other.

\subsection{Role-Conditioned Search Intent Distillation}

Directly using the experience repository \(\mathcal{R}_t\) as context requires the proposer \(\mathcal{P}\) to perform evidence selection and search planning simultaneously, implicitly deciding which evidence to exploit, avoid, or revisit, as well as which component to modify and how to modify it. This can destabilize search directions across iterations. Even with relevance-based retrieval or history compression, the resulting context remains primarily descriptive of previous trials, leaving its implications for subsequent evolution to \(\mathcal{P}\)'s implicit reasoning. To address this issue, we introduce \textit{Role-Conditioned Search Intent Distillation}, an evidence-conditioned search policy that distills calibrated evidence into an explicit search intent specifying the target component and corresponding modification direction before \(\mathcal{P}\) generates \(h_{t+1}\).

Specifically, after running $h_t$ in iteration $t$, DREvo extracts current failure patterns $\mathcal{F}_t$ from the trial log $\tau_t$ and directly retrieves historical evidence associated with the same failure categories from $\mathcal{R}_t$. We denote the retrieved evidence set by $\mathcal{Z}_t$. For each evidence unit $e\in\mathcal{Z}_t$, DREvo obtains its current polarity $\rho_e\in\{+1,-1\}$ and its weight $w_e\in[0,1]$ calibrated by State-Dependent Evidence Recalibration.
DREvo first partitions the retrieved evidence into three decision sets according to its polarity and weight. Given a confidence threshold $\theta=0.5$ (midpoint of $w_e\in[0,1]$), the evidence partition is defined as:
\begin{equation}
\begin{aligned}
\mathcal{X}_t
&=
\{e\in\mathcal{Z}_t\mid w_e>\theta,\rho_e=+1\},\\
\mathcal{A}_t
&=
\{e\in\mathcal{Z}_t\mid w_e>\theta,\rho_e=-1\},\\
\mathcal{T}_t
&=
\{e\in\mathcal{Z}_t\mid w_e\leq\theta\}.
\end{aligned}
\end{equation}

Here, $\mathcal{X}_t$, $\mathcal{A}_t$, and $\mathcal{T}_t$ contain reliable positive evidence, reliable negative evidence, and low-confidence evidence, respectively.
Based on these decision sets, Role-Conditioned Search Intent Distillation defines four search roles: \textsc{Exploit}, \textsc{Avoid}, \textsc{Retest}, and \textsc{Explore}. Each role assigns a distinct operational meaning to the retrieved evidence and prescribes a corresponding search direction for subsequent evolution.

The \textsc{Exploit} role treats reliable positive evidence from $\mathcal{X}_t$ as binding guidance, requiring $\mathcal{P}$ to follow the corresponding direction and inherit or refine the validated modifications. The \textsc{Avoid} role uses reliable negative evidence from $\mathcal{A}_t$ to avoid known harmful modifications while leaving the remaining search space open. The \textsc{Retest} role presents low-confidence evidence from $\mathcal{T}_t$ as non-binding references, allowing $\mathcal{P}$ to independently decide whether to follow or avoid the corresponding directions. Unlike these evidence-driven roles, \textsc{Explore} imposes no evidence-derived constraints and allows $\mathcal{P}$ to freely determine the component and modification direction based on the current failure patterns.

After evidence partition, DREvo adopts a role activation strategy that prioritizes high-confidence positive evidence, as shown in Algorithm~\ref{alg:steering_policy}. Let $a_t$ denote the activated search role, and let $\mathrm{failRetest}_{t-1}$ indicate whether the previous iteration was guided by \textsc{Retest} but yielded no performance improvement. \textsc{Retest} is used to reassess low-confidence evidence under the current harness state. If a \textsc{Retest}-guided iteration is unsuccessful, it switches to \textsc{Explore} in the next iteration to avoid repeated attempts along uncertain directions.
\begin{algorithm}[t]
\caption{Role Activation}
\label{alg:steering_policy}
\begin{algorithmic}[1]
\REQUIRE Evidence sets $(\mathcal{X}_t,\mathcal{A}_t,\mathcal{T}_t)$,
retrieved evidence $\mathcal{Z}_t$, and previous \textit{Retest-failure} flag
$\mathrm{failRetest}_{t-1}$
\ENSURE Activated search role $a_t$
\IF{$\mathcal{X}_t\neq\varnothing$}
    \STATE $a_t\leftarrow\textsc{Exploit}$
\ELSIF{$\mathcal{A}_t\neq\varnothing$}
    \STATE $a_t\leftarrow\textsc{Avoid}$
\ELSIF{$\mathcal{Z}_t=\varnothing\ \lor\
\mathrm{failRetest}_{t-1}$}
    \STATE $a_t\leftarrow\textsc{Explore}$
\ELSIF{$\mathcal{T}_t\neq\varnothing$}
    \STATE $a_t\leftarrow\textsc{Retest}$
\ELSE
    \STATE $a_t\leftarrow\textsc{Explore}$
\ENDIF
\RETURN $a_t$
\end{algorithmic}
\end{algorithm}
Then, DREvo distills the current failure patterns and role-associated evidence into an explicit search intent:
\begin{equation}
s_t=
G\!\left(
a_t,\mathcal{F}_t,\mathcal{Z}_t^{a_t}
\right),
\qquad
h_{t+1}\sim\mathcal{P}(\cdot\mid h_t,s_t),
\end{equation}
where $\mathcal{Z}_t^{a_t}$ denotes the subset of retrieved evidence associated with the activated role $a_t$, and $G$ instantiates the corresponding role semantics to distill the current failure patterns and role-associated evidence into a target component and corresponding modification direction.


\begin{table*}[t]
\centering
\small
\setlength{\tabcolsep}{6pt}
\resizebox{\textwidth}{!}{%
\begin{tabular}{l|cccccc@{\hspace{1em}}|l|cccc}
\toprule

\multicolumn{7}{c|}{Domain Reasoning} &
\multicolumn{5}{c}{Agentic Tasks} \\
\cmidrule(lr){1-7}\cmidrule(lr){8-12}

\multirow[c]{2}{*}{Method}
& \multicolumn{2}{c}{USPTO}
& \multicolumn{2}{c}{S2D}
& \multicolumn{2}{c}{Law}
& \multirow[c]{2}{*}{Method}
& \multicolumn{2}{c}{Terminal-Bench 2.0}
& \multicolumn{2}{c}{SWE-Bench Verified} \\
\cmidrule(lr){2-3}\cmidrule(lr){4-5}\cmidrule(lr){6-7}
\cmidrule(lr){9-10}\cmidrule(lr){11-12}

& Acc. & $\Delta$
& Acc. & $\Delta$
& Acc. & $\Delta$
&
& Acc. & $\Delta$
& Acc. & $\Delta$ \\
\midrule

Naive
& 12.0 & 13.0
& 63.2 & 26.0
& 7.0 & 42.0
& OpenHands
& 31.5 & 11.2
& 61.8 & 5.8 \\

Few-Shot (8)
& 14.0 & 11.0
& 72.2 & 17.0
& 21.0 & 28.0
& Terminus-KIRA
& 25.8 & 16.9
& 41.4 & 26.2 \\

Few-Shot (32)
& 13.0 & 12.0
& 67.9 & 21.3
& 21.0 & 28.0
& Terminus-2
& \underline{34.8} & 7.9
& 49.6 & 18.0 \\

Few-Shot (all)
& 15.0 & 10.0
& 78.3 & 10.9
& 29.0 & 20.0
& Mini-SWE-Agent
& 23.6 & 19.1
& \underline{66.8} & 0.8 \\

MCE
& 14.0 & 11.0
& 83.0 & 6.2
& 23.0 & 26.0
& Meta-harness
& 29.2 & 13.5
& 38.6 & 29.0 \\

ACE
& \underline{16.0} & 9.0
& 77.8 & 11.4
& 29.0 & 20.0
& Claude Code
& 24.7 & 18.0
& 55.2 & 12.4 \\

Meta-harness
& 14.0 & 11.0
& \underline{86.8} & 2.4
& \underline{45.0} & 4.0
& AHE
& -- & --
& 61.8 & 5.8 \\
\midrule

\textbf{DREvo (Ours)}
& \textbf{25.0} & --
& \textbf{89.2} & --
& \textbf{49.0} & --
& \textbf{DREvo (Ours)}
& \textbf{42.7} & --
& \textbf{67.6} & -- \\

\bottomrule
\end{tabular}%
}

\caption{
Overall performance (\%). $\Delta$ denotes the absolute improvement of DREvo over each baseline. Bold and underlined values indicate the best and second-best accuracy.
}
\label{tab:overall_results}
\end{table*}

\begin{table*}[t]
\centering
\footnotesize
\setlength{\tabcolsep}{5.5pt}
\resizebox{\textwidth}{!}{%
\begin{tabular}{l|cccc|ccccccccc}
\toprule

\multirow[c]{2}{*}{Method}
& \multicolumn{4}{c|}{Terminal-Bench 2.0}
& \multicolumn{9}{c}{SWE-Bench Verified} \\
\cmidrule(lr){2-5}\cmidrule(lr){6-14}

& Easy & Med. & Hard & Full
& Django & SymPy & Sphinx & MPL
& Sklearn & PyData & Astropy & Others & Full \\
\midrule

\# Tasks
& 4 & 55 & 30 & 89
& 231 & 75 & 44 & 34
& 32 & 22 & 22 & 40 & 500 \\
\midrule

OpenHands
& \underline{50.0} & \underline{41.8} & 10.0
& 31.5
& 66.2 & \underline{65.3}
& \underline{50.0} & \textbf{67.6}
& 62.5 & 36.4
& \textbf{50.0} & \underline{57.5}
& 61.8 \\

Terminus-KIRA
& \textbf{75.0} & 34.5 & 3.3
& 25.8
& 48.1 & 33.3 & 31.8 & 29.4
& 62.5 & 27.3 & 36.4 & 32.5
& 41.4 \\

Terminus-2
& \textbf{75.0} & \underline{41.8} & \underline{16.7}
& \underline{34.8}
& 55.4 & 46.7 & 36.4 & 47.1
& 68.8 & 54.5 & 27.3 & 32.5
& 49.6 \\

Mini-SWE-Agent
& 25.0 & 30.9 & 10.0
& 23.6
& \textbf{73.6} & \textbf{69.3}
& \textbf{54.5} & \underline{58.8}
& \underline{71.9} & \textbf{86.4}
& 36.4 & 45.0
& \underline{66.8} \\

Meta-harness
& \underline{50.0} & 34.5 & \underline{16.7}
& 29.2
& 41.6 & 37.3 & 29.5 & 32.4
& 62.5 & 36.4 & 22.7 & 30.0
& 38.6 \\

Claude Code
& \textbf{75.0} & 29.1 & 10.0
& 24.7
& 64.5 & 53.3 & 43.2 & 32.4
& 68.8 & \underline{63.6} & 27.3 & 37.5
& 55.2 \\
\midrule

\textbf{DREvo (Ours)}
& \textbf{75.0} & \textbf{49.1} & \textbf{26.7}
& \textbf{42.7}
& \underline{68.4} & \underline{65.3}
& \textbf{54.5} & \textbf{67.6}
& \textbf{90.6} & \textbf{86.4}
& \underline{45.5} & \textbf{65.0}
& \textbf{67.6} \\

\bottomrule
\end{tabular}%
}

\caption{
Fine-grained success rates (\%) by task difficulty and repository.
Others include Pytest, Pylint, Requests, Seaborn, and Flask.
Bold and underlined values denote the best and second-best accuracy,
respectively. Matplotlib is abbreviated as MPL.
}
\label{tab:fine_grained_results}
\end{table*}


\section{Experiments} 

\subsection{Experimental Setups}

\noindent
\textbf{Tasks and Datasets.}
(1) Domain Reasoning.
We consider USPTO for chemical reaction prediction~\cite{schneider2016s}, Symptom2Disease (S2D) for medical diagnosis~\cite{lee2026meta}, and Law for legal reasoning~\cite{fei2024lawbench}. These tasks evaluate context organization, example selection, and reasoning strategies. Each dataset is divided into a validation set for harness evolution and a test set for the final evaluation of the best-performing evolved harness.
(2) Agentic Tasks.
We choose Terminal-Bench 2.0 (TB2) for terminal interaction~\cite{merrill2026terminal} and SWE-Bench Verified for code repair~\cite{jimenez2024swe}. These benchmarks evaluate tool use, state management, error recovery, and long-horizon execution. Following existing work~\cite{lee2026meta}, we perform harness self-evolution and evaluation on the original agentic benchmark suites without additional splits. The proposer receives only structured harness-level evidence extracted from logs, excluding reference solutions and task-specific target outputs. Consequently, the reported gains reflect benchmark-specific harness adaptation rather than generalizable improvements on unseen tasks.

\noindent
\textbf{Baselines.}
(1) Domain Reasoning.
We compare against the non-evolutionary baselines Naive (no\_memory), Few-Shot (8), Few-Shot (32), and Few-Shot (all), as well as the evolution-based baselines MCE, ACE, and Meta-harness.
(2) Agentic Tasks.
We compare against the non-evolutionary agents OpenHands, Terminus-KIRA, Terminus-2, Mini-SWE-Agent, and Claude Code (v2.1.150), together with the evolution-based baselines AHE and Meta-harness.

\noindent
\textbf{Models and Evolution Setting.}
All evolution-based methods use Claude Opus 4.6 \textit{(1M-token context window)} as the proposer. All methods are evaluated with GPT-OSS-120B for domain reasoning tasks and DeepSeek-V4-Flash \textit{(no thinking)} for agentic tasks. Evolution-based methods run for 20 iterations on domain reasoning tasks and 10 iterations on agentic tasks.
We initialize the harness by combining
the no\_memory and Few-Shot (all) configurations for domain reasoning tasks and using Terminus-KIRA for agentic tasks.

\subsection{Experimental Results}
In this section, we present the main experiments, including overall and fine-grained performance comparisons, evolution-trajectory analysis, ablation studies, and robustness analysis under component-state drift. 

\noindent
\textbf{Overall Performance.}
Table~\ref{tab:overall_results} compares DREvo with representative manually designed harnesses and harness self-evolution methods. DREvo achieves the highest point estimate on all five benchmarks under the evaluated configuration and a limited evolutionary budget. On domain reasoning, it reaches 25.0\% on USPTO, 89.2\% on S2D, and 49.0\% on Law, exceeding the second-best methods by 9.0, 2.4, and 4.0 percentage points, respectively. On agentic tasks, DREvo also outperforms manually engineered task-specific harnesses and several existing harness-evolution methods. Notably, the agentic evolution starts from the terminal-oriented Terminus-KIRA rather than a coding-specific harness. On SWE-Bench Verified, DREvo evolves the terminal-oriented initial harness into a competitive code-repair harness through direct evolution under the target environment, without requiring manual task-specific engineering. Overall, consistent gains in reasoning and agentic tasks demonstrate the effectiveness of DREvo in diverse task domains and execution environments.

\noindent
\textbf{Fine-Grained Analysis.}
Table~\ref{tab:fine_grained_results} provides a fine-grained comparison across task difficulties and software repositories. On TB2, DREvo matches the best result on Easy tasks and achieves the highest success rates on Medium and Hard tasks. Its gains increase from 7.3 points on Medium tasks to 10.0 points on Hard tasks, indicating greater effectiveness as execution complexity increases.
On SWE-Bench Verified, DREvo ranks first or ties for first on five of the eight repository groups and ranks second on the remaining three. It achieves the best results on Scikit-learn and Others, while matching the best methods on Sphinx, Matplotlib, and PyData. This broad coverage shows that its overall improvement is not driven by a single repository or software ecosystem.

\begin{figure}
    \centering
    \includegraphics[width=\linewidth]{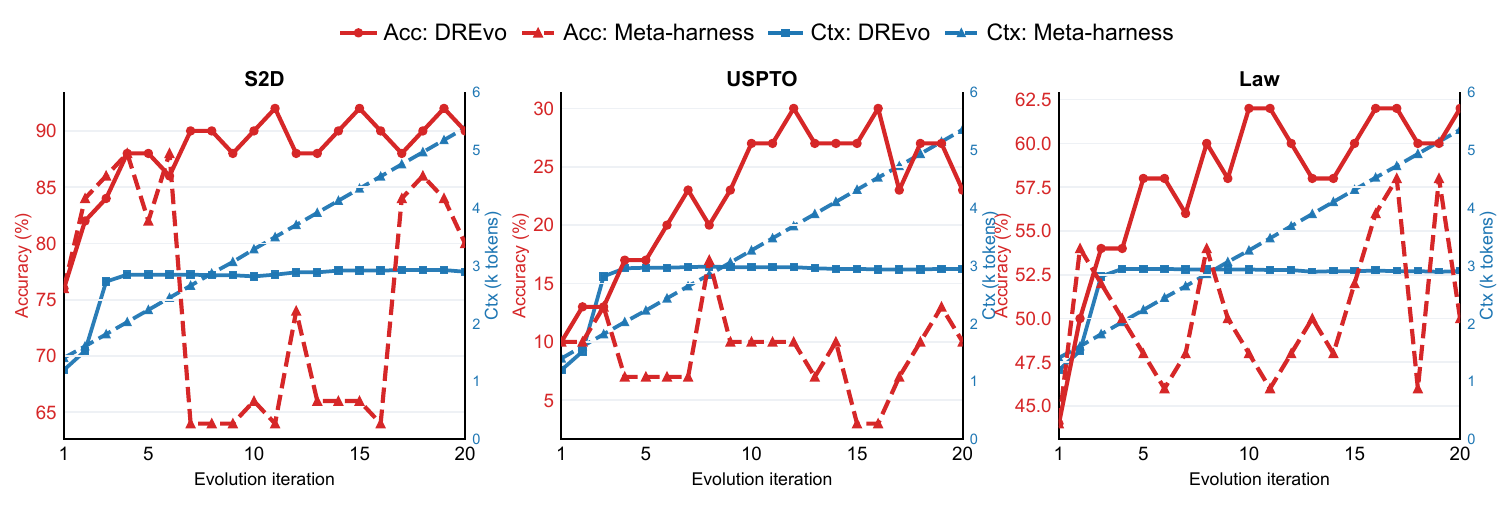}
    \caption{
    Validation set accuracy (Acc.) and context (Ctx.) size during harness self-evolution for DREvo and Meta-harness (compressed-history) on domain reasoning tasks.}
    \label{fig:acc}
\end{figure}

\begin{figure}
    \centering
    \includegraphics[width=\linewidth]{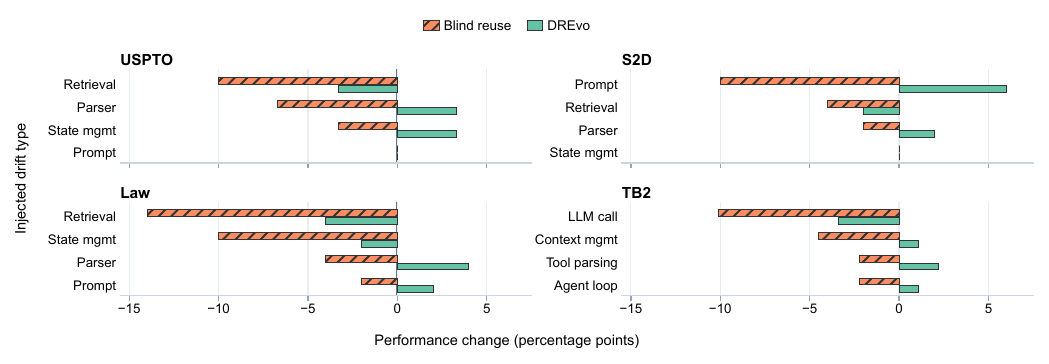}
    \caption{Performance changes under controlled component state drift during harness evolution. Drift perturbs the target component while keeping historical evidence fixed. Blind Reuse applies evidence without state calibration.}
    \label{fig:state}
\end{figure}

\noindent
\textbf{Effectiveness Analysis.}
\begin{table}[t]
\centering
\resizebox{\columnwidth}{!}{
\begin{tabular}{ccc|ccc|c|ccc}
\toprule

\multicolumn{3}{c}{Components}
& \multicolumn{3}{c}{Domain Reasoning}
& \multicolumn{4}{c}{Terminal-Bench 2.0} \\
\cmidrule(lr){1-3}
\cmidrule(lr){4-6}
\cmidrule(lr){7-10}

A & R & D
& USPTO & S2D & Law
& Full & Easy & Med. & Hard \\
\midrule

$\times$ & $\times$ & $\times$
& 12.0 & 78.8 & 34.0
& 29.2 & 50.0 & 34.5 & 16.7 \\

$\checkmark$ & $\times$ & $\times$
& 20.0 & 81.1 & 37.0
& 29.2 & \underline{75.0} & 34.5 & 13.3 \\

$\checkmark$ & $\checkmark$ & $\times$
& \underline{23.0} & \underline{86.3} & 39.0
& 35.9 & \underline{75.0}
& \underline{40.0} & \underline{23.3} \\

$\checkmark$ & $\times$ & $\checkmark$
& 18.0 & 84.0 & \underline{42.0}
& \underline{37.1} & \textbf{100.0}
& 38.2 & \textbf{26.7} \\
\midrule

$\checkmark$ & $\checkmark$ & $\checkmark$
& \textbf{25.0} & \textbf{89.2} & \textbf{49.0}
& \textbf{42.7} & \underline{75.0}
& \textbf{49.1} & \textbf{26.7} \\

\bottomrule
\end{tabular}
}

\caption{
Ablation results (\%). A, R, and D denote Function-Level Evidence Anchoring,
State-Dependent Evidence Recalibration, and Role-Conditioned Search Intent Distillation.
}
\label{tab:component_ablation}
\end{table}
Figure~\ref{fig:acc} compares validation accuracy and proposer context size during harness self-evolution. DREvo improves in early iterations and follows smoother accuracy trajectories with fewer large oscillations than Meta-harness. Despite the overhead introduced by its structured search specification, DREvo keeps the proposer context bounded at approximately 2.9K tokens. By contrast, Meta-harness exhibits substantial accuracy fluctuations as its context grows beyond 5.3K tokens. Importantly, DREvo continues to improve without expanding the proposer context, indicating that its gains do not simply result from exposing the proposer to more history or from context compression alone. Together, these patterns support the effectiveness of recalibrating evidence under the current harness state and distilling it into an explicit, targeted search intent.
Figure~\ref{fig:state} evaluates robustness to component-state drift by prompting Claude Opus 4.6 to randomly modify one component of the final evolved harness. Across 16 settings covering four drift types and four benchmarks, blind evidence reuse decreases performance by 5.3 points on average, whereas DREvo improves it by 0.6 points. DREvo maintains or improves performance in 11 settings, compared with only 2 under blind reuse. The largest gap occurs under prompt drift on S2D, where the performance change improves from \(-10.0\) to \(+6.0\) points. Although degradation remains under retrieval and LLM-call drift, DREvo consistently mitigates the impact of outdated evidence across all four benchmarks.

\noindent
\textbf{Ablation Study.}
Table~\ref{tab:component_ablation} presents the ablation results. A alone improves history-driven evolution by 8.0, 2.3, and 3.0 points on the three reasoning tasks, indicating that function-level anchoring makes historical feedback more attributable and reusable, although it leaves TB2 unchanged. Adding R improves all four results, including a 6.7-point gain on TB2, confirming the value of reassessing evidence validity as the harness evolves. D particularly benefits harder agentic tasks but is less consistent without R: A+D achieves 100.0\% on Easy and 26.7\% on Hard, yet trails the full model on every aggregate result. Combining A, R, and D achieves the strongest aggregate performance across all benchmarks and exceeds A+D by 5.6 points on TB2. These results demonstrate their complementary roles: A anchors feedback to attributable modifications, R identifies evidence that remains reliable under the current state, and D distills the calibrated evidence into actionable search directions.

\section{Conclusion and Future Work}
In this paper, we investigated the non-stationary search behavior exhibited by history-driven harness self-evolution under both full-history and compressed-history reuse of accumulated trial experience. We identified two underlying limitations: \textit{historical evidence may become invalid as the harness state evolves}, and \textit{it does not directly specify which component to modify or how to modify it}. To address these limitations, we proposed \textbf{DREvo}, which organizes historical trial experience into function-anchored evidence, recalibrates its validity under the current harness state, and distills calibrated evidence into an explicit role-conditioned search intent. DREvo improves performance across five benchmarks and yields smoother evolution trajectories under limited evolution budgets. Future work will further investigate cross-component evidence calibration and adaptive role-selection policies for longer-horizon harness self-evolution.

\bibliography{aaai2027}

@article{lee2026meta,
  title={Meta-harness: End-to-end optimization of model harnesses},
  author={Lee, Yoonho and Nair, Roshen and Zhang, Qizheng and Lee, Kangwook and Khattab, Omar and Finn, Chelsea},
  journal={arXiv preprint arXiv:2603.28052},
  year={2026}
}

@article{yang2024swe,
  title={Swe-agent: Agent-computer interfaces enable automated software engineering},
  author={Yang, John and Jimenez, Carlos and Wettig, Alexander and Lieret, Kilian and Yao, Shunyu and Narasimhan, Karthik and Press, Ofir},
  journal={Advances in Neural Information Processing Systems},
  volume={37},
  pages={50528--50652},
  year={2024}
}

@inproceedings{wang2025openhands,
  title={Openhands: An open platform for ai software developers as generalist agents},
  author={Wang, Xingyao and Li, Boxuan and Song, Yufan and Xu, Frank F and Tang, Xiangru and Zhuge, Mingchen and Pan, Jiayi and Song, Yueqi and Li, Bowen and Singh, Jaskirat and others},
  booktitle={International Conference on Learning Representations},
  volume={2025},
  pages={65882--65919},
  year={2025}
}

@article{zhang2025agentic,
  title={Agentic context engineering: Evolving contexts for self-improving language models},
  author={Zhang, Qizheng and Hu, Changran and Upasani, Shubhangi and Ma, Boyuan and Hong, Fenglu and Kamanuru, Vamsidhar and Rainton, Jay and Wu, Chen and Ji, Mengmeng and Li, Hanchen and others},
  journal={arXiv preprint arXiv:2510.04618},
  year={2025}
}

@article{ye2026meta,
  title={Meta Context Engineering via Agentic Skill Evolution},
  author={Ye, Haoran and He, Xuning and Arak, Vincent and Dong, Haonan and Song, Guojie},
  journal={arXiv preprint arXiv:2601.21557},
  year={2026}
}

@article{li2024embodied,
  title={Embodied agent interface: Benchmarking llms for embodied decision making},
  author={Li, Manling and Zhao, Shiyu and Wang, Qineng and Wang, Kangrui and Zhou, Yu and Srivastava, Sanjana and Gokmen, Cem and Lee, Tony and Li, Li E and Zhang, Ruohan and others},
  journal={Advances in Neural Information Processing Systems},
  volume={37},
  pages={100428--100534},
  year={2024}
}

@inproceedings{liu2026toolscope,
  title={Toolscope: Enhancing llm agent tool use through tool merging and context-aware filtering},
  author={Liu, Marianne Menglin and Garcia, Daniel and Parllaku, Fjona and Upadhyay, Vikas and Shah, Fahad and Roth, Dan},
  booktitle={Proceedings of the 64th Annual Meeting of the Association for Computational Linguistics (Volume 1: Long Papers)},
  pages={34095--34119},
  year={2026}
}

@inproceedings{guo2025dior,
  title={Dior: Adaptive cognitive detection and contextual retrieval optimization for dynamic retrieval-augmented generation},
  author={Guo, Hanghui and Zhu, Jia and Di, Shimin and Shi, Weijie and Chen, Zhangze and Xu, Jiajie},
  booktitle={Proceedings of the 63rd Annual Meeting of the Association for Computational Linguistics (Volume 1: Long Papers)},
  pages={2953--2975},
  year={2025}
}

@inproceedings{zhang2025aflow,
  title={Aflow: Automating agentic workflow generation},
  author={Zhang, Jiayi and Xiang, Jinyu and Yu, Zhaoyang and Teng, Fengwei and Chen, Xionghui and Chen, Jiaqi and Zhuge, Mingchen and Cheng, Xin and Hong, Sirui and Wang, Jinlin and others},
  booktitle={International Conference on Learning Representations},
  volume={2025},
  pages={34040--34077},
  year={2025}
}

@inproceedings{qiao2025benchmarking,
  title={Benchmarking agentic workflow generation},
  author={Qiao, Shuofei and Fang, Runnan and Qiu, Zhisong and Wang, Xiaobin and Zhang, Ningyu and Jiang, Yong and Xie, Pengjun and Huang, Fei and Chen, Huajun},
  booktitle={International Conference on Learning Representations},
  volume={2025},
  pages={69679--69703},
  year={2025}
}

@article{lin2026agentic,
  title={Agentic harness engineering: Observability-driven automatic evolution of coding-agent harnesses},
  author={Lin, Jiahang and Liu, Shichun and Pan, Chengjun and Lin, Lizhi and Dou, Shihan and Xi, Zhiheng and Huang, Xuanjing and Yan, Hang and Han, Zhenhua and Gui, Tao and others},
  journal={arXiv preprint arXiv:2604.25850},
  year={2026}
}

@article{zhang2026self,
  title={Self-Harness: Harnesses That Improve Themselves},
  author={Zhang, Hangfan and Zhang, Shao and Li, Kangcong and Zhang, Chen and Chen, Yang and Zhang, Yiqun and Bai, Lei and Hu, Shuyue},
  journal={arXiv preprint arXiv:2606.09498},
  year={2026}
}

@inproceedings{hu2025automated,
  title={Automated design of agentic systems},
  author={Hu, Shengran and Lu, Cong and Clune, Jeff},
  booktitle={International Conference on Learning Representations},
  volume={2025},
  pages={21344--21377},
  year={2025}
}

@inproceedings{yang2024large,
  title={Large language models as optimizers},
  author={Yang, Chengrun and Wang, Xuezhi and Lu, Yifeng and Liu, Hanxiao and Le, Quoc V and Zhou, Denny and Chen, Xinyun},
  booktitle={International Conference on Learning Representations},
  volume={2024},
  pages={12028--12068},
  year={2024}
}

@article{agrawal2025gepa,
  title={Gepa: Reflective prompt evolution can outperform reinforcement learning},
  author={Agrawal, Lakshya A and Tan, Shangyin and Soylu, Dilara and Ziems, Noah and Khare, Rishi and Opsahl-Ong, Krista and Singhvi, Arnav and Shandilya, Herumb and Ryan, Michael J and Jiang, Meng and others},
  journal={arXiv preprint arXiv:2507.19457},
  year={2025}
}

@article{ning2026code,
  title={Code as agent harness},
  author={Ning, Xuying and Tieu, Katherine and Fu, Dongqi and Wei, Tianxin and Li, Zihao and Bei, Yuanchen and Zou, Jiaru and Ai, Mengting and Liu, Zhining and Li, Ting-Wei and others},
  journal={arXiv preprint arXiv:2605.18747},
  year={2026}
}

@article{meng2026agent,
  title={Agent harness for large language model agents: A survey},
  author={Meng, Qianyu and Wang, Yanan and Chen, Liyi and Wang, Qimeng and Lu, Chengqiang and Wu, Wei and Gao, Yan and Wu, Yi and Hu, Yao},
  year={2026},
  publisher={Preprints}
}

@article{wang2026openclaw,
  title={Openclaw-rl: Train any agent simply by talking},
  author={Wang, Yinjie and Chen, Xuyang and Jin, Xiaolong and Wang, Mengdi and Yang, Ling},
  journal={arXiv preprint arXiv:2603.10165},
  year={2026}
}

@article{merrill2026terminal,
  title={Terminal-bench: Benchmarking agents on hard, realistic tasks in command line interfaces},
  author={Merrill, Mike A and Shaw, Alexander G and Carlini, Nicholas and Li, Boxuan and Raj, Harsh and Bercovich, Ivan and Shi, Lin and Shin, Jeong Yeon and Walshe, Thomas and Buchanan, E Kelly and others},
  journal={arXiv preprint arXiv:2601.11868},
  year={2026}
}

@misc{krafton2026terminus,
  title={Terminus-KIRA: Boosting frontier model performance on Terminal-Bench with minimal harness},
  author={KRAFTON, AI and Robotics, Ludo},
  year={2026}
}

@inproceedings{jimenez2024swe,
  title={Swe-bench: Can language models resolve real-world github issues?},
  author={Jimenez, Carlos E and Yang, John and Wettig, Alexander and Yao, Shunyu and Pei, Kexin and Press, Ofir and Narasimhan, Karthik},
  booktitle={International Conference on Learning Representations},
  volume={2024},
  pages={54107--54157},
  year={2024}
}

@inproceedings{liang2026rsda,
  title={RSDA: Restoring Stale Data Affinity via Dynamic Renovation Strategy for Mitigating Data Scarcity},
  author={Liang, Yidan and Zhu, Jia and Shi, Weijie and Guo, Hanghui and Cui, Yue and Shen, Jiawei and Ma, Guoqing and Liu, Jingjiang and Niu, Qingyu and Wang, Yilin and others},
  booktitle={Proceedings of the 64th Annual Meeting of the Association for Computational Linguistics (Volume 1: Long Papers)},
  pages={8280--8309},
  year={2026}
}

@inproceedings{shen2026acr,
  title={ACR: Adaptive Context Refactoring via Context Refactoring Operators for Multi-Turn Dialogue},
  author={Shen, Jiawei and Zhu, Jia and Guo, Hanghui and Shi, Weijie and Cui, Yue and Niu, Qingyu and Ma, Guoqing and Liu, Jingjiang and Liang, Yidan and Wang, Yilin and others},
  booktitle={Findings of the Association for Computational Linguistics: ACL 2026},
  pages={3149--3167},
  year={2026}
}

@article{lou2026autoharness,
  title={Autoharness: improving llm agents by automatically synthesizing a code harness},
  author={Lou, Xinghua and L{\'a}zaro-Gredilla, Miguel and Dedieu, Antoine and Wendelken, Carter and Lehrach, Wolfgang and Murphy, Kevin P},
  journal={arXiv preprint arXiv:2603.03329},
  year={2026}
}

@article{chen2026unlocking,
  title={Unlocking the Codex harness: how we built the app server},
  author={Chen, Celia},
  journal={OpenAI engineering note},
  year={2026}
}

@article{rajasekaran2026harness,
  title={Harness design for long-running application development},
  author={Rajasekaran, Prithvi},
  journal={Anthropic Engineering Blog. URL: https://www. anthropic. co m/engineering/harness-design-long-running-apps. accessed},
  pages={05--07},
  year={2026}
}

@article{lopopolo2026harness,
  title={Harness engineering: leveraging codex in an agent-first world},
  author={Lopopolo, Ryan},
  journal={OpenAI Engineering Blog},
  year={2026}
}

@inproceedings{fei2024lawbench,
  title={Lawbench: Benchmarking legal knowledge of large language models},
  author={Fei, Zhiwei and Shen, Xiaoyu and Zhu, Dawei and Zhou, Fengzhe and Han, Zhuo and Huang, Alan and Zhang, Songyang and Chen, Kai and Yin, Zhixin and Shen, Zongwen and others},
  booktitle={Proceedings of the 2024 conference on empirical methods in natural language processing},
  pages={7933--7962},
  year={2024}
}

@article{schneider2016s,
  title={What’s what: The (nearly) definitive guide to reaction role assignment},
  author={Schneider, Nadine and Stiefl, Nikolaus and Landrum, Gregory A},
  journal={Journal of chemical information and modeling},
  volume={56},
  number={12},
  pages={2336--2346},
  year={2016},
  publisher={ACS Publications}
}


\end{document}